\documentclass[a4paper,dvips 2t,superscriptaddress,showkeys]{revtex4} 
\usepackage{epsfig,graphics,graphicx,amsmath,amssymb}

\begin{document}
\title{\bf\bf NLO radiative correction to the Casimir energy in Lorentz-violating scalar field theory}

\author{Amirhosein Mojavezi}\email{amojavezi98@gmail.com}
\affiliation{Department of Physics, Ferdowsi University of Mashhad, 91775-1436 Mashhad, I.R. Iran}
\author{Reza Moazzemi }\email{r.moazzemi@qom.ac.ir}\affiliation{Department of Physics, University of Qom, Ghadir Blvd., Qom 371614-6611, I.R. Iran} \author{Mohammad Ebrahim Zomorrodian}\affiliation{Department of Physics, Ferdowsi University of Mashhad, 91775-1436 Mashhad, I.R. Iran}
\begin{abstract}
Violation of the Lorentz symmetry has  important effects on physical quantities including field  propagators. Therefore, in addition to the leading order, the sub-leading order of quantities may be modified. In this paper, we calculate the next to leading (NLO) radiative corrections to the Casimir energy in the presence of two perfectly conducting parallel plates for $\phi^4$ theory with a Lorentz-breaking extension. We do the renormalization and investigate these NLO corrections for three distinct directions of the Lorentz violation;  temporal direction, parallel and perpendicular to the plates.
\end{abstract} 
\maketitle

\section{Introduction}
The Casimir effect which is a physical manifestation of changes in the quantum vacuum fluctuations for different configurations, was discovered by H. B. G. Casimir in 1948 \cite{casimir1948influence}. He showed the existence of this effect as an attractive force between two infinite parallel uncharged perfectly conducting plates in vacuum (for a general review on the Casimir effect, see Refs. \cite{bordag2001new,milton2001casimir}). Sparnaay \cite{sparnaay1958measurements} and Arnold et al \cite{arnold1979influence}  experimentally observed the Casimir force for such a configuration. Also, the other measurements, with  greatly improved precisions, have been done for various geometries \cite{garcia2012casimir,kim2010surface,lamoreaux2010reanalysis}.

In addition to the leading Casimir energy, the next to leading order (NLO)  radiative corrections to this effect is an exciting subject of discussion. The first endeavors to calculate the leading radiative corrections to the Casimir energy were reported in \cite{bordag1983quantum}. Also, many works on the radiative corrections to the Casimir energy for various cases exist in the literature (see for instance \cite{bordag1998radiative,moazzemi2007dirichlet}). In the case of a real massive scalar field, NLO correction to the Casimir energy has been computed in \cite{bordag2001new,barone2004radiative}. We have also calculated one loop radiative corrections to the Casimir energy in \cite{moazzemi2016one}.

In original quantum field theory (QFT), the Lorentz symmetry is preserved. However, there are some theories which present models with Lorentz symmetry violation (for example\cite{ferrari2013hovrava,ulion2015casimir}). Naturally, Lorentz symmetry violation arises from, for example, existence of space-time anisotropy \cite{colladay1997cpt,colladay1998lorentz} or non-commutativity \cite{carroll2001noncommutative,bertolami2003noncommutative} or a spacetime varying coupling constant \cite{kostelecky2003spacetime,bertolami1997lorentz}. Investigations of Casimir effect with Lorentz-breaking symmetry for QED theory have been done (see please \cite{frank2006casimir,martin2016casimir,martin2017local}). It has also been studied recently for a real massive scalar field  in \cite{cruz2017casimir}. 

In this paper we calculate the NLO correction to the Casimir energy  in an interacting scalar field theory, $\lambda \phi^4$,  with a Lorentz violating term. Our configuration is two perfectly conducting parallel plates. We work within the renormalized perturbation theory, therefore we need to reconsider the renormalization for this theory. Naturally, the counterterms needed for renormalization, are modified due to the existence of new Lorentz violating terms in the Lagrangian. 

To take the physical result and resolve infinities problem, we use a well-known approach called \textit{Boyer method}\cite{boyer1968quantum}; also is known as \textit{Box Renormalization Scheme} (BRS). This method uses a completely physical approach by enclosing the whole system in a box of volume $V=L^3$ which finally may tend to infinity in such a way that difference between the zero point energies of two different configurations is calculated. It removes all ambiguities associated with appearance of the infinities without resorting to any other schemes such as analytic continuation approach. It is notable that, in BRS the substraction precedure in calculation of Casimir energy takes place in two physical configuration with similar nature, which is another advantage of BRS. 

We organized our paper as follows:

We introduce our model for Lorentz-breaking symmetry of the theory in section \ref{2} . We shall see that energy-momentum tensor and Klein-Gordon (KG) equation is modified. In section \ref{3}, we survey renormalization of the related theory within a Lorentz-braeking case. In section \ref{4} we calculate the NLO radiative correction to the Casimir energy for $\phi^4$ theory with Lorentz-breaking symmetry. We note that  at this stage we consider the existence of Lorentz-symmetry parameter in two cases: 1. time-like (TL), and 2. space-like (SL). Finally, in last section we state our conclusions. 

\section{The Lorentz-Breaking $\phi^4$ theory}\label{2}
\subsection{The Model}
In this section, we present the Lorentz symmetry breaking for a scalar field theory due to an anisotropy of space-time. We do this by insertion an additional term in the KG Lagrangian density
\begin{equation}\label{phi}
  {\cal L}(x)
  =\frac{1}{2}[\partial_{\mu}\phi]^{2}+\frac{1}{2} c(u.\partial\phi)^2  -\frac{1}{2}m_0^{2}\phi^{2},
\end{equation}
where $m_{0}$ is the bare mass and the dimensionless parameter $c$, which  is  much smaller than one, manifests the  Lorentz symmetry breaking of the system by a coupling between the derivative of the scalar field $\phi$ and a constant four-vector $u^\mu$. Adding a self-interaction term to Eq. \eqref{phi} we get
\begin{equation}\label{phi2}
{\cal L}(x)
=\frac{1}{2}(\partial_{\mu}\phi)^{2}+\frac{1}{2} c(u.\partial\phi)^2  -\frac{1}{2}m_0^{2}\phi^{2}-\frac{\lambda_0}{4!}\phi^4,
\end{equation}
where $\lambda_0$ is our bare coupling. The equation of motion for Lagrangian \eqref{phi}  reads as
\begin{eqnarray}\label{MKG}
\left[\square+c (u\cdot\partial)^2-m_0^2\right]\phi(x)=0.
\end{eqnarray}
It is obvious that this modified KG equation, for  $c= 0$ reverts to the original KG equation of motion with the following dispersion relation: 
\begin{equation}\label{DR}
 \omega_n ^2=\left \vert\mathbf{k}^{\bot}\right \vert ^2+k_n^2+m_0^2.
\end{equation}
The violation of Lorentz symmetry has vital consequences such as modification of dispersion relation which directly affects the propagator of the field. We consider this effect in three different cases. In the first case we assume that the Lorentz violation is in the time direction. The second and third are the SL Lorentz violations in the directions parallel (pl-SL) and perpendicular (pr-SL) to the plates. 

\subsection{Propagator in Bounded Space}
To calculate radiative corrections to any physical quantity, including  Casimir energy, we need to know the  exact form of propagator. In this subsection we first derive the propagator, suitable for Casimir effect problem, in the context of standard quantum field theory (without any Lorentz-violating term). Our configuration is two parallel plates located at $z=\pm a/2$ perpendicular to $z$-axis with a separation $a$. We suppose the fields satisfy  Dirichlet boundary conditions (DBCs) on the plates,
\begin{equation}
\phi \left(x\right)
\bigg|_{z=\pm a/2}=0.
\end{equation}
Being $d$ the dimension of space-time, the field $\phi$ is defined with quantized modes as 

\begin{eqnarray}
   \phi(x)&=&\int\frac{d^{d-2}\mathbf{k}^{\bot}}{(2\pi)^{d-1}}
   \sum_{n=1}^{\infty}\left(\frac{1}{a\omega_{n}}\right)^{1/2}
   \nonumber\\
   &&\hspace{-.2cm} \times\Bigg\{e^{-i(\omega_{n}t-\mathbf{k^{\bot}}\textbf{.}
   \mathbf{x}^{\bot})}\sin\left[k_{n}(z+\frac{a}{2})\right]\textbf{a}_{n} +e^{i(\omega_{n}t-\mathbf{k^{\bot}}\textbf{.}
   \mathbf{x}^{\bot})}\sin\left[k_{n}(z+\frac{a}{2})\right]
   \textbf{a}_{n}^{\dag}\Bigg\},
\end{eqnarray}
where $\mathbf{k}^{\bot}$ and $k_{n}=\frac{n\pi}{a}$ denote the momenta parallel and perpendicular to the plates, respectively.

Here,
$\textbf{a}_{n}^{\dag}$ and $\textbf{a}_{n}$ are creation and annihilation operators, respectively, with the following commutation relations:
\[[\textbf{a}_{n},\textbf{a}^{\dag}
_{n^{_\prime}}]=\delta_{n,n^{_\prime}},\quad
[\textbf{a}_{n},\textbf{a}_{n^{_\prime}}]=[\textbf{a}^{\dag}_{n},
\textbf{a}^{\dag}_{n^{_\prime}}]=0,\]
 and $\textbf{a}|0\rangle=0$
defines the vacuum state in the presence of boundary conditions.
One may easily find Feynman Green's function of the KG equation as
\begin{eqnarray}
  &&\hspace{-.7cm}G_F(x,x')=i\frac{2}a\int\frac{d\omega}{2\pi}\int\frac{d^{d-2} {\bf
k^{\bot}}}{(2\pi)^{d-2}} \sum_n
\frac{e^{-i\omega(t-t')}e^{-i{\bf
k^{\bot}}.({\bf{x}}^{\bot}-{\bf{x'}}^{\bot})}
  \sin\left[k_{n}(z+\frac{a}{2})\right]
\sin\left[k_{n}(z'+\frac{a}{2})\right]}{\omega^2 -{
k^{\bot^2}}-k_{n}^2-m_0^2+i\epsilon}.\nonumber\\
\end{eqnarray}
We then find Euclidean Green's function by the following definitions:

\[\omega_E=-i\omega \hspace{.5cm}; \hspace{1cm} {\bf
k^{\bot}_E}={\mathbf{ k}^{\bot}}, \]
which finally  leads to (we need only $G_F(x,x)$ in our calculations)
\begin{eqnarray}\label{Green}
  \hspace{-.7cm}G_F(x,x)=\frac{2}a\int\frac{d\omega_E}{2\pi}\int\frac{d^{d-2} \mathbf{k}^{\bot}_E}{(2\pi)^{d-2}} \sum_n
\frac{
  \sin^2\left[k_{n}(z+\frac{a}{2})\right]
}{\omega^2_E +{{k}^{\bot}_E}^2+k_{n}^2+m_0^2+i\epsilon}.
\end{eqnarray}

\subsubsection{TL vector case}
Choosing the four-vector to be TL, $u^\mu=(1,0,0,0)$, the second term in Eq. \eqref{MKG} becomes $c\, \partial_0^2$. Hence, the dispersion relation \eqref{DR} takes the form  
\begin{eqnarray}
 (1+c)\omega^{2}_{n}={
 	k^{\bot}}^2+k_{n}^2+m_0^2.
\end{eqnarray}
Therefore, we can find the propagator for this case by replacing
 $\omega^2\to(1+c)\omega^2$ in Eq. (\ref{Green})
\begin{eqnarray}
  G_F(x,x)=\frac{2}a\int\frac{d\omega_E}{2\pi}\int\frac{d^{d-2} \mathbf{k}^{\bot}_E}{(2\pi)^{d-2}} \sum_n
  \frac{
  	\sin^2\left[k_{n}(z+\frac{a}{2})\right]
  }{(1+c)\omega^2_E +{{k}^{\bot}_E}^2+k_{n}^2+m_0^2+i\epsilon}
\end{eqnarray}
Changing variable $\omega'=\sqrt{1+c} \hspace{.1cm}\omega_E$    , we obtain
\begin{eqnarray}
\hspace{-.7cm}G_F(x,x)=\frac{2}{a(1+c)^{1/2}}\int\frac{d^{d-1} {
k}}{(2\pi)^{d-1}} \sum_n
\frac{
  \sin^2\left[k_{n}(z+\frac{a}{2})\right]
}{k^2+k_{n}^2+m_0^2+i\epsilon}.
\end{eqnarray}
where $\mathbf{k}=(\omega',{\mathbf k}^\bot_E)$ . Performing the angular integration, finally we have 
\begin{eqnarray}\label{Gtime}
  G_F(x,x)&=&\frac{2}{a(1+c)^{1/2}}\Omega_{d-1}\int \frac{dk k^{d-2}}{(2\pi)^{d-1}}  \sum_n
\frac{
  \sin^2\left[k_{n}(z+\frac{a}{2})\right]
}{{k^2+k_n^2+m_0^2+i\epsilon}}.\nonumber\\&=&\frac{4}{a(1+c)^{1/2}(4\pi)^{\frac{d-1}{2}}\Gamma(\frac{d-1}{2})}\int dk k^{d-2} \sum_n
\frac{
	\sin^2\left[k_{n}(z+\frac{a}{2})\right]
}{{k^2+k_n^2+m_0^2+i\epsilon}}
\end{eqnarray}
where the solid angle $\Omega_d=\dfrac{2\pi^{d/2}}{\Gamma(d/2)}$, with $\Gamma(x)$ being the Gamma function, corresponds to the area of a unit sphere in $d$ dimensions.
\subsubsection{SL vector case}
In SL case we choose three distinct directions for four-vector $u^\mu$; $u^\mu=(0,1,0,0)$, $u^\mu=(0,0,1,0)$ and $u^\mu=(0,0,0,1)$. In this case  the Lorentz-breaking term in  \eqref{MKG} is $-c\partial_i^2$ with $i=x,y$ or $z$. There is no difference between the physics of  the first two vectors (pl-SL case), which are parallel to the plates, and the dispersion relations for both cases are also the same. Choosing $u^\mu=(0,0,1,0)$ for instance, Eq. \eqref{DR} becomes
\begin{eqnarray}
\omega^{2}_{n}=k_x^2+(1-c)
	k_y^2+k_{n}^2+m_0^2.
\end{eqnarray}
Changing the variables $\mathbf{k}=(\omega,{\mathbf k'}^\bot_E)$ with $k'_y=\sqrt{1-c} \ k_y$, in a similar manner to the TL case, the Green's function is derived as:
\begin{eqnarray}\label{SL}
  \hspace{-.7cm}G_F(x,x)=\frac{4}{(1-c)^{1/2} a (4\pi)^{\frac{d-1}{2}}\Gamma(\frac{d-1}{2})}\int dk k^{d-2} \sum_n
\frac{
  \sin^2\left[k_{n}(z+\frac{a}{2})\right]
}{{k^2+k_n^2+m_0^2+i\epsilon}}.
\end{eqnarray}
Now, for the last case (pl-SL),  $u^\mu=(0,0,0,1)$ is normal to the plates and  Eq. \eqref{DR} becomes
\begin{equation}
  \omega^{2}_{n}={
  	k^{\bot}}^2+(1-c)k_{n}^2+m_0^2.
\end{equation}
In this case, the Euclidean Feynman propagator is derived as
\begin{eqnarray}
  G_F(x,x)&=&\frac{4}{a (4\pi)^{\frac{d-1}{2}}\Gamma(\frac{d-1}{2})}\int dk k^{d-2} \sum_n
\frac{
  \sin^2\left[k_{n}(z+\frac{a}{2})\right]
}{{k^2+(1-c) k_n^2+m_0^2+i\epsilon}}.\nonumber\\
&=&\frac{4}{a (4\pi)^{\frac{d-1}{2}}\Gamma(\frac{d-1}{2})}\int dk k^{d-2} \sum_n
\frac{
	\sin^2\left[k_{n}(z+\frac{a}{2})\right]
}{{k^2+ {k'}_n^2+m_0^2+i\epsilon}}.\label{Glike}
\end{eqnarray}
where $k'_n=n \pi/a'$ with $a'=a/\sqrt{1-c}$.

For the future use, in the case of free space without plates, we note that the propagator for a Lorentz symmetry breaking theory becomes 
\begin{eqnarray}\label{fgf}
G_F(x,x)
&=&\frac{1}{(1\pm c)^{\frac{1}{2}}}\int\frac{d^dk}{(2\pi)^d}\frac{i}{k^2-m_0^2}
\\\nonumber&=&\frac{1}{(1\pm c)^{\frac{1}{2}}(4\pi)^\frac{d}{2}}\frac{\Gamma(1-\frac{d}{2})}{(m_0^2)^{1-\frac{d}{2}}}.
\end{eqnarray}
where $+$ ($-$) is used for TL (SL) vector case. 
\section{Renormalization up to order $\lambda$}\label{3}

At the level of quantum corrections, all unphysical quantities such as $m_0$ and $\lambda_0$ need to be renormalized. Therefore, we need to do a renormalization procedure to extract the physical $m$ and $\lambda$ from the bare parameters $m_0$ and $\lambda_0$ (see \cite{peskin1995introduction}). Here, we work within the standard renormalized perturbation theory. In the Lagrangian \eqref{phi2}, after rescaling the fields by a field strength renormalization $Z$, namely  $\phi=Z^{\frac{1}{2}} \phi_r$ we have

\begin{eqnarray}\label{Ren}
   &&{\cal L}=\frac{1}{2}(\partial_{\mu}\phi_{r})^{2}+\frac{1}{2} c(u.\partial\phi_r)^2-\frac{1}{2}m^{2}
   \phi_{r}^{2}-\frac{\lambda}{4!}\phi_{r}^{4}\qquad\nonumber\\&&\hspace{2cm}+
   \frac{1}{2}\delta_{Z}(\partial_{\mu}\phi_{r})^{2}
   -\frac{1}{2}\delta_{m}\phi_{r}^{2}-\frac{\delta_{\lambda}}{4!}\phi_{r}^{4},
\end{eqnarray}
where $\delta_m=m_0^2 Z-m^2$, $\delta_\lambda=\lambda_0 Z^2-\lambda$ and $\delta_Z=Z-1$ are the counterterms. Then, we have
two new Feynman rules from the above Lagrangian
\begin{eqnarray}
\raisebox{-3mm}{\includegraphics[scale=.3]{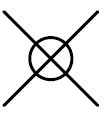}}&=&-i\delta_{\lambda}\nonumber
\\
\raisebox{-1mm}{\includegraphics[scale=.3]{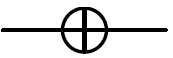}}&=&i[\left(c\pm 1\right)p^\mu p_\mu\delta_{Z}-\delta_{m}],
\end{eqnarray}
where $+$ ($-$) along with $\mu=0$ ($\mu=i$) are used for TL (SL) vector case (for more details see \cite{ferrero2011renormalization}). The counterterms are totally fixed by two renormalization conditions:

\begin{eqnarray}
\raisebox{-5mm}{\includegraphics[scale=.15]{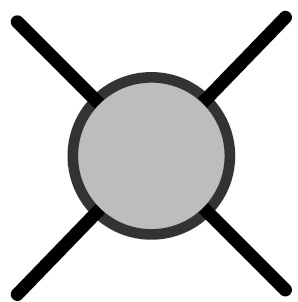}}&=&-i\lambda \qquad (s=4m^2, t=u=0) \\\nonumber
\raisebox{-3.1mm}{\includegraphics[scale=0.15]{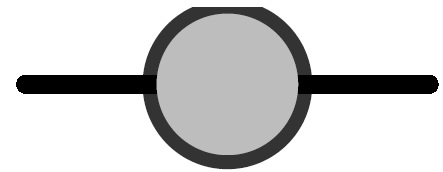}}&=&\frac{i}{p^2-m^2}+(\text{terms regular  at $p^2=m^2$}).
\end{eqnarray}
From the first renormalization condition it is obvious that $\delta_\lambda=O(\lambda^2)$. The second renormalization condition which gives the physical mass $m$, up to order $\lambda$, can be written as 
\begin{eqnarray}\label{delta}
	0&=&\raisebox{-.5mm}{\includegraphics[scale=.2]{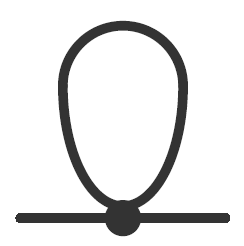}}+
	\raisebox{-1.3mm}{\includegraphics[scale=0.3]{5}}\nonumber\\
	&=&-\frac{1}{2}\dfrac{i\lambda}{(1\pm c)^{1/2}(4\pi)^\frac{d}{2}}\dfrac{\Gamma(1-\frac{d}{2})}{(m^2)^{1-\frac{d}{2}}}+i[ (c\pm1)p^\mu p_\mu\delta_{Z}-\delta_{m}]
\end{eqnarray}
where we have used Eq. \eqref{fgf}. Therefore,  $\delta_Z$ up to order $\cal O (\lambda)$ is zero, and
\begin{equation}\label{deltam}
\delta_m=-\frac{\lambda}{2(1\pm c)^{1/2}(4\pi)^\frac{d}{2}}\frac{\Gamma(1-\frac{d}{2})}{(m^2)^{1-\frac{d}{2}}}.
\end{equation}

\section{Radiative correction to the Casimir energy }\label{4}
\begin{figure}[th]
	\begin{center} \includegraphics[width=7cm]{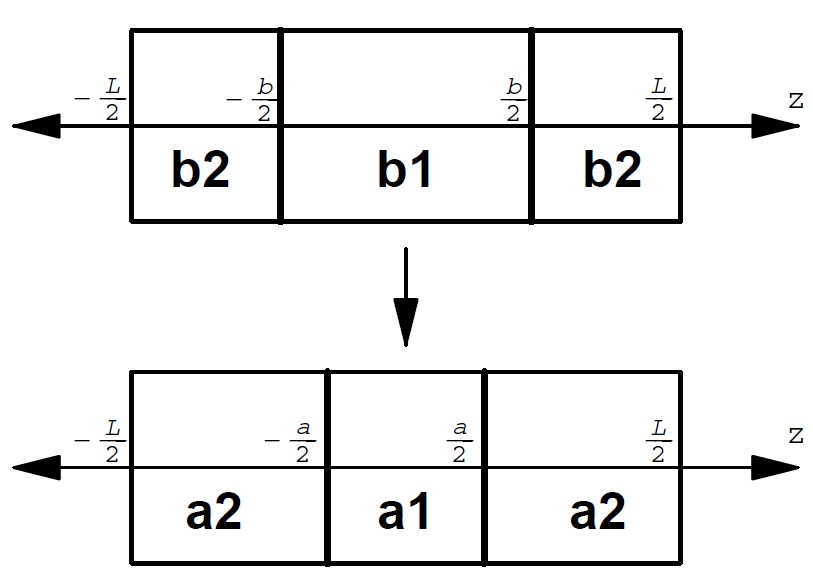}\caption{\label{fig.1} {\small The labels a1, {\em etc.} denote the appropriate sections in each configuration separated by the plates.}}
		\label{geometry}
	\end{center}
\end{figure}

In order to calculate the radiative correction to the Casimir energy we use box renormalization scheme (BRS) \cite{boyer1968quantum}. In this approach, we first compare the energies in two various configurations: when the plates are at $\pm a/2$ as compared to $\pm b/2$. We confine each configuration in a box with edges are located at $\pm L/2$ in all directions (see figure \ref{geometry}). 
Now, the Casimir energy is defined as
\begin{equation}\label{Cas1}
E_{\mbox{\tiny \mbox{\tiny Cas.}}}=\lim_{b/a\rightarrow\infty}\left[\lim_{L/b\rightarrow\infty}
\left(E_{a}-E_{b}\right)\right],
\end{equation}
where,
\begin{equation}\label{e3}
E_{a}=E_{a1}+2E_{a2},\quad
E_{b}=E_{b1}+2E_{b2}.
\end{equation} 
The radiative corrections to the zero point energy in the (for example)  a1 part, i.e. $z \in [\frac{-a}{2},\frac{a}{2}]$,  are
\begin{eqnarray}\label{e21:vacc-pol}
\Delta E_{a1}= E^{(1)}_{a1}+ E^{(2)}_{a1}+\dots= \int_{V}
d^3{\bf x}\langle\Omega|{\cal H}_{_I}|\Omega\rangle\qquad\nonumber\\=i\int_{V} d^3{\bf x}\left(\ \frac{1}{2}
\raisebox{-1mm}{\includegraphics[scale=.3]{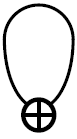}}+\frac{1}{8}
\raisebox{-7mm}{\includegraphics[scale=.3]{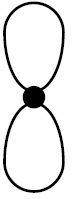}} \ +\frac{1}{8}\raisebox{-7mm}{\includegraphics[scale=.3]{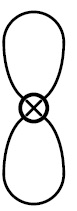}}+\dots
\right),
\end{eqnarray}
where $|\Omega\rangle$ is the vacuum state in the presence of interaction. Up to  order ${}\lambda$
we have
\begin{eqnarray}\label{Rcas}
   E^{(1)}_{a_1}&=& i\int_{V} d^3{\bf x}\left(\ \frac{1}{2}
   \raisebox{-1mm}{\includegraphics[scale=.3]{2}}+\frac{1}{8}
   \raisebox{-7mm}{\includegraphics[scale=.3]{1}} \ \right)\hspace{3cm}\nonumber\\&=&i\int_{V}
   d^3{\bf x}\left[-\frac{i}{2}
   \delta_{m}G_{a1}(x,x)-\frac{i\lambda}{8}G^{2}_{a1}(x,x)\right]
   ,
\end{eqnarray}
where $G_{a_1}(x,x)$ is the propagator of the real scalar field in region a1 (we drop the subscript `F' for simplicity). 
\subsection{TL \& pl-SL vector cases}
 To calculate the  first term in  Eq. (\ref{Rcas}), $E^{(1),F}_{a_1}$, using Eqs. (\ref{Gtime}) and \eqref{deltam}  and carrying out the spatial integration, one obtains the correction to the vacuum energy in region a1, up to ${\cal O}(\lambda)$,  as:
\begin{eqnarray}
 E^{(1),F}_{a_1}=\frac{1}{2}\int_{V}
\delta_{m}G_{a1}(x,x)d^3{\bf x}=-\frac{\lambda \sqrt{\pi}  \Gamma(1-\frac{d}{2})L^{d-2}}{2 (1\pm c) (4\pi)^d (m^2)^{1-\frac{d}{2}} \Gamma(\frac{d-1}{2})}\int_0^\infty dk k^{d-2} \sum_n \frac{1}{k^2+k_{a1,n}^2+m^2},
\end{eqnarray}
where $k_{a1,n}=\frac{n \pi}{a}$. Integrating over momentum $k$ yields
\begin{eqnarray}\label{Cas2}
 \frac{1}{2}\int_{V}
 \delta_{m}G_{a_1}(x,x)d^3{\bf x}=\frac{\lambda \pi^{\frac{3}{2}}   \Gamma(1-\frac{d}{2}) \sec(\frac{d \pi}{2})L^{d-2}}{4(1 \pm c) (4\pi)^d (m^2)^{1-\frac{d}{2}} \Gamma(\frac{d-1}{2})}\sum_n {\omega}_{a_1,n}^{d-3},
\end{eqnarray}
where ${\omega}_{a_1,n}=(m^{2}+k_{a_1,n}^{2})^{1/2}$. This is one of the four terms (related to the $a1$ region) that contribute to the NLO radiative correction for Casimir energy Eq. (\ref{Cas1}). To derive the Casimir energy from Eq. \eqref{Cas2},  we apply Abel-Plana summation formula \cite{saharian2007generalized},
\begin{equation}\label{Abel}
    \sum_{n=1}^{\infty}g(n)=-\frac{g(0)}{2}+\int_{0}^{\infty}g(x)dx
    +i\int_{0}^{\infty}\frac{g(it)-g(-it)}{e^{2\pi t}-1}dt,
\end{equation}
with,
\begin{eqnarray}
   g(n)={\omega}_{a1,n}^{d-3}+{\omega}_{a2,n}^{d-3}-{\omega}_{b1,n}^{d-3}-{\omega}_{b2,n}^{d-3}.
   \end{eqnarray}
We note that the $g(0)$ term vanishes. Also the second term on the right hand side of Eq. \eqref{Abel},  with respect to suitable changing of variables in the four integrals below, vanishes:
\begin{eqnarray}
   && \frac{a}
   {\pi}\int^{\infty}_0dk'\left(m^2+k'^2\right)^{(d-3)/2}+2\frac{L-a}{2\pi}
   \int^{\infty}_0dk'\left(m^2+k'^2\right)^{(d-3)/2}\nonumber\\
 &&\quad-\frac{b}{\pi}\int^{\infty}_0dk'
   \left(m^2+k'^2\right)^{(d-3)/2}-2\frac{L-b}{2\pi}\int^{\infty}_0dk'
   \left(m^2+k'^2\right)^{(d-3)/2}=0.
\end{eqnarray}
Finally, we calculate branch-cut terms in Eq. \eqref{Abel}. Assuming $f(x)=[x^2-(\frac{am}{\pi})^2]^{(d-3)/2}$ we have
\begin{eqnarray}\label{Branch}
B(a)=i \int_0^\infty\frac{f(it)-f(-it)}{e^{2\pi t}-1} dt&=&-2\left(\frac{\pi}{a}\right)^{d-3} \int_{\frac{am}{\pi}}^\infty\nonumber \frac{[t^2-(\frac{am}{\pi})^2]^{(d-3)/2}}{e^{2\pi t}-1}\\
&=&-\sum_{j=1}^\infty\frac{2 K_{\frac{d-2}{2}}(2amj)\Gamma(\frac{d-1}{2})}{a^{d-3}\pi^{3/2}(\frac{am}{j})^{\frac{2-d}{2}}},
\end{eqnarray}
where $K_n(x)$ is the modified Bessel function of order $n$. To calculate the integral, we have used the identity
\begin{equation}
\frac{1}{e^{2 \pi t}-1}=\sum_{j=1}^\infty e^{- 2 \pi j t}.
\end{equation}
Therefore, we obtain

\begin{equation}\label{Ea}
 E^{(1),F}_{a_1}=-\frac{\lambda \pi^{\frac{3}{2}}   \Gamma(1-\frac{d}{2}) \sec(\frac{d \pi}{2})L^{2}}{4(1 \pm c) a^3 (4\pi)^d (\tilde{m}^2)^{1-\frac{d}{2}} \Gamma(\frac{d-1}{2})}\sum_{j=1}^\infty\frac{2 K_{\frac{d-2}{2}}(2\tilde{m}j)\Gamma(\frac{d-1}{2})}{\pi^{3/2}(\frac{\tilde{m}}{j})^{\frac{2-d}{2}}},
\end{equation}
where $\tilde{m}=ma$ is a dimensionless parameter. Then, according to Eq. \eqref{Cas1} the contribution of 
the Eq. \eqref{Cas2} to Casimir energy is  
\begin{eqnarray}\label{Cas3}
E^{(1),F}_{_{\mbox{\tiny Cas.}}}=\lim_{b/a\rightarrow\infty}\left[\lim_{L/b\rightarrow\infty}
\left( E^{(1),F}_{a_1}- E^{(1),F}_{b_1}+ E^{(1),F}_{a_2}- E^{(1),F}_{b_2}\right)\right].
\end{eqnarray}
Taking the limits, only the first term  survives. Finally, we take the limit $d\to4$,
\begin{eqnarray}\label{RCAS2}
E^{(1),F}_{_{\mbox{\tiny Cas.}}}=-\sum_{j=1}^\infty\frac{\lambda \tilde{m}^3 L^2}{512 (1 \pm c) a^3 \pi^4} \frac{1}{j}\left[K_1(2\tilde{m}j)\left(\ln(\frac{\tilde{m}^3}{16\pi^2j})+\gamma-1\right)+K_1^{'}(2\tilde{m}j)\right],
\end{eqnarray}
 where $K'_q(x)=\frac{\partial}{\partial q}K_q(x)$ and $\gamma$ is the Euler-Mascheroni number.

The contribution of the second term in Eq. (\ref{Rcas}) to the Casimir energy, $ E^{(1),S}_{a_1}$,  without Lorentz violating terms, have been calculated in Ref. \cite{moazzemi2007dirichlet} using BRS:
 \begin{eqnarray}
 E^{(1),S}_{a_1}&=&\frac{\lambda}{8}\int_{V}
 G^{2}_{a1}(x,x)d^3{\bf x}\nonumber
  \\\label{Branch2}
 &&\to E^{(1),S}_{_{\mbox{\tiny Cas.}}}={-\lambda L^2}
 \frac{B(a)}{128\pi^2}\left[\frac{B(a)}{a}
 -\frac{m}{a}+\frac{m^2}{\pi}(\ln2+1/2)
 \right]\label{Rcas.}\qquad\mbox{(no Lorentz violation)}
 \\
 &&\hspace{1.5cm}={-\lambda L^2}
 \sum_{j=1}^{\infty}\frac{m}{128\pi^3}\frac{K_{1}(2amj)}{j}\left[\frac{m}{\pi a}
 \sum_{j'=1}^{\infty}\frac{K_{1}(2amj')}{j'}+\frac{m}{a}-\frac{m^2}{\pi}(\ln2+1/2)
 \right].\label{Rcas..}
 \end{eqnarray}
 But, when we have a TL (pl-SL) Lorentz  breaking term, an extra  factor $\frac{1}{\sqrt{1+c}}$ ($\frac{1}{\sqrt{1-c}}$), as we see in Eq. \eqref{Gtime} (Eq. \eqref{SL}), is multiplied to the propagator. Therefore to derive the Casimir energy contribution we only need to multiply the factor  $\frac{1}{1+c}$ ($\frac{1}{1-c}$) to Eq. \eqref{Rcas..}.
Accordingly, using Eqs. \eqref{Rcas..} and \eqref{RCAS2},  we can write NLO radiative correction to the Casimir enegy as
\begin{eqnarray}\label{final2}
\hspace{-.7cm}E^{(1)}_{_{\mbox{\tiny Cas.}}}=E^{(1),F}_{_{\mbox{\tiny Cas.}}}+E^{(1),S}_{_{\mbox{\tiny Cas.}}}=\frac{-\lambda L^2}{(1\pm c)}
\sum_{j=1}^{\infty}\bigg\{\frac{m}{128\pi^3}\frac{K_{1}(2amj)}{j}\bigg[\frac{m}{\pi a}
\sum_{j'=1}^{\infty}\frac{K_{1}(2amj')}{j'}+\frac{m}{a}-\frac{m^2}{\pi}(\ln2+1/2)
\bigg]\\\nonumber
-\frac{m^3 }{512 \pi^4} \frac{1}{j}\bigg[K_1(2amj)\left(\ln(\frac{a^3m^3}{16\pi^2j})+\gamma-1\right)+K_1^{'}(2amj)\bigg]\bigg\}.
\end{eqnarray}
 From this result it is obvious that the influence of the Lorentz-symmetry breaking parameter appears only in a factor.
 
  Two special limits are interesting to calculate; the large mass $ma\gg1$, and small mass  $m\to0$ limits:
\begin{equation}
 \left\{
   \begin{array}{ll}
      E^{(1)}_{_{\mbox{\tiny Cas.}}}\quad{\buildrel {am\gg1 } \over
 \longrightarrow }\quad
\displaystyle\frac{ 3 \ 
L^2 }{1024 \pi^{7/2}}\frac{\lambda}{(1 \pm c) a^3} (am)^{5/2}
 \  \ln(am) \ e^{-2am},
     &  \\
     \raisebox{-9mm}{$E^{(1)}_{_{\mbox{\tiny Cas.}}}
   \quad {\buildrel {m\to0 } \over
 \longrightarrow }\quad \displaystyle -\frac{L^2\lambda }{512\pi^4 (1\pm c)a^3}
   \left(\sum_{j=1}^{\infty}\frac{1}{j^2}\right)^2=-\frac{L^2\lambda }{18432 (1\pm c) a^3},$}
     &
   \end{array}\right.
\end{equation}
with $+$ ($-$) for TL (pl-SL) case.
\subsection{pr-SL vector case}
For the pr-SL vector case, $u^\mu=(0,0,0,1)$, we do not need to do  new calculation. In this case, applying Eq. \eqref{Glike} leads us to the following expression for Eq. \eqref{Cas2}:
\begin{eqnarray}\label{Cas2p}
\frac{1}{2}\int_{V}
\delta_{m}G_{a_1}(x,x)d^3{\bf x}=\frac{\lambda \pi^{\frac{3}{2}}   \Gamma(1-\frac{d}{2}) \sec(\frac{d \pi}{2})L^{d-2}}{4(1-c)^{1/2} (4\pi)^d (m^2)^{1-\frac{d}{2}} \Gamma(\frac{d-1}{2})}\sum_n {\omega'}_{a_1,n}^{d-3}
\end{eqnarray}
where ${\omega'}_{a_1,n}=(m^{2}+{k'}_{a_1,n}^{2})^{1/2}$. Therefore, the Eq. \eqref{Branch} becomes
\begin{eqnarray}
B(a')&=&-\sum_{j=1}^\infty\frac{2 K_{\frac{d-2}{2}}(2a'mj)\Gamma(\frac{d-1}{2})}{a'^{d-3}\pi^{3/2}(\frac{a'm}{j})^{\frac{2-d}{2}}}\\
&=&-\sum_{j=1}^\infty\frac{2 K_{\frac{d-2}{2}}\left(\frac{2amj}{\sqrt{1-c}}\right)\Gamma(\frac{d-1}{2})}{(1-c)^{\frac{3-d}2}a^{d-3} \pi^{3/2}\left(\frac{am}{j\sqrt{1-c}}\right)^{\frac{2-d}{2}}},
\end{eqnarray}
and hence, we get 
\begin{equation}
E^{(1),F}_{a_1}=-\frac{\lambda \pi^{\frac{3}{2}}   \Gamma(1-\frac{d}{2}) \sec(\frac{d \pi}{2})L^{2}}{4(1-c)^{\frac{6-d}{4}} {a}^3 (4\pi)^d (\tilde{m}^2)^{1-\frac{d}{2}} \Gamma(\frac{d-1}{2})}\sum_{j=1}^\infty\frac{2 K_{\frac{d-2}{2}}\left(\frac{2\tilde{m}j}{\sqrt{1-c}}\right)\Gamma(\frac{d-1}{2})}{\pi^{3/2}\left(\frac{\tilde{m}}{j}\right)^{\frac{2-d}{2}}}.
\end{equation}
Now, we use the above equation to compute Eq. \eqref{Cas3}, and take the limit $d\to4$, to get
\begin{eqnarray}\label{final22}
\hspace{-.7cm}E^{(1),F}_{_{\mbox{\tiny Cas.}}}=\frac{\lambda L^2 m^3}{(1-c)^{1/2}512 \pi^4} \sum_{j=1}^{\infty}\frac{1}{j}\bigg[K_1(\frac{2amj}{\sqrt{1-c}})\left(\ln(\frac{a^3m^3}{16\pi^2j})+\ln(1-c)+\gamma-1\right)+K_1^{'}(\frac{2amj}{\sqrt{1-c}})\bigg].
\end{eqnarray}
 
 Similary, for the second term in Eq. \eqref{Rcas}, now the Eq. \eqref{Rcas.} becomes
\begin{eqnarray}
 E^{(1),S}_{_{\mbox{\tiny Cas.}}}&=&-\frac{\lambda L^2}{(1-c)^{1/2}}
\frac{B(a')}{128\pi^2}\left[\frac{B(a')}{a}
-\frac{m}{a}+\frac{m^2}{\pi}(\ln2+1/2)
\right]\nonumber\\
&{\buildrel {d\to4 } \over
	= }&-\frac{\lambda L^2}{(1-c)^{1/2}}
\sum_{j=1}^{\infty}\bigg\{\frac{m}{128\pi^3}\frac{1}{j}K_{1}(\frac{2amj}{\sqrt{1-c}})\bigg[\frac{m}{\pi a}
\sum_{j'=1}^{\infty}\frac{1}{j'}K_{1}(\frac{2amj'}{\sqrt{1-c}})+\frac{m}{a}-\frac{m^2}{\pi}(\ln2+1/2)
\bigg]
\end{eqnarray}
 Therefore the result for the radiative correction of Casimir energy for the pr-SL vector case can be written as 
\begin{eqnarray}
\hspace{-.7cm}E^{(1)}_{_{\mbox{\tiny Cas.}}}&=&E^{(1),F}_{_{\mbox{\tiny Cas.}}}+E^{(1),S}_{_{\mbox{\tiny Cas.}}}\nonumber\\
&=&-\frac{\lambda L^2}{(1-c)^{1/2}} \sum_{j=1}^{\infty}\bigg\{\frac{m^3}{512 \pi^4}\frac{1}{j}\bigg[K_1(\frac{2amj}{\sqrt{1-c}})\left(\ln(\frac{a^3m^3}{16\pi^2j})+\ln(1-c)+\gamma-1\right)+K_1^{'}(\frac{2amj}{\sqrt{1-c}})\bigg]\nonumber\\
&&+\frac{m}{128\pi^3}\frac{1}{j}K_{1}(\frac{2amj}{\sqrt{1-c}})\bigg[\frac{m}{\pi a}
\sum_{j'=1}^{\infty}\frac{1}{j'}K_{1}(\frac{2amj'}{\sqrt{1-c}})+\frac{m}{a}-\frac{m^2}{\pi}(\ln2+1/2)
\bigg]\bigg\}
\end{eqnarray}
\begin{figure}[h]
	\begin{center} \includegraphics[width=16cm]{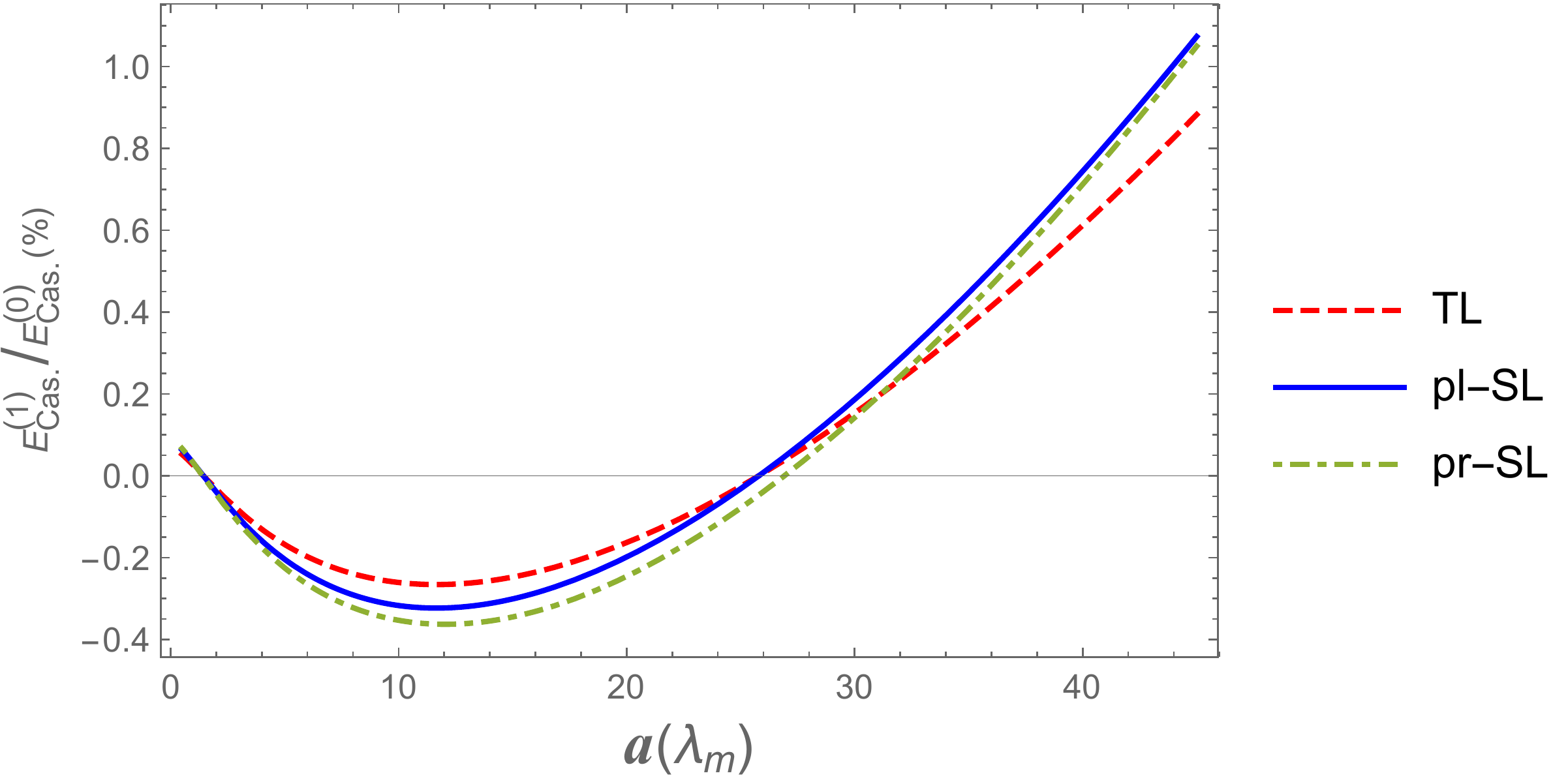} \caption{ {\small The ratio between the first order radiative corrections and leading terms, $E^{(1)}_{_{\mbox{\tiny Cas.}}}/E^{(0)}_{_{\mbox{\tiny Cas.}}}$, in terms of plates separation $a$, for $c=0.1$, $\lambda=0.1$ and $ m=1$; $\lambda_m$ is the Compton wavelength of the scalar field.}}
		\label{fig1}
	\end{center}
\end{figure}

\begin{figure}[h]
	\begin{center} \includegraphics[width=16cm]{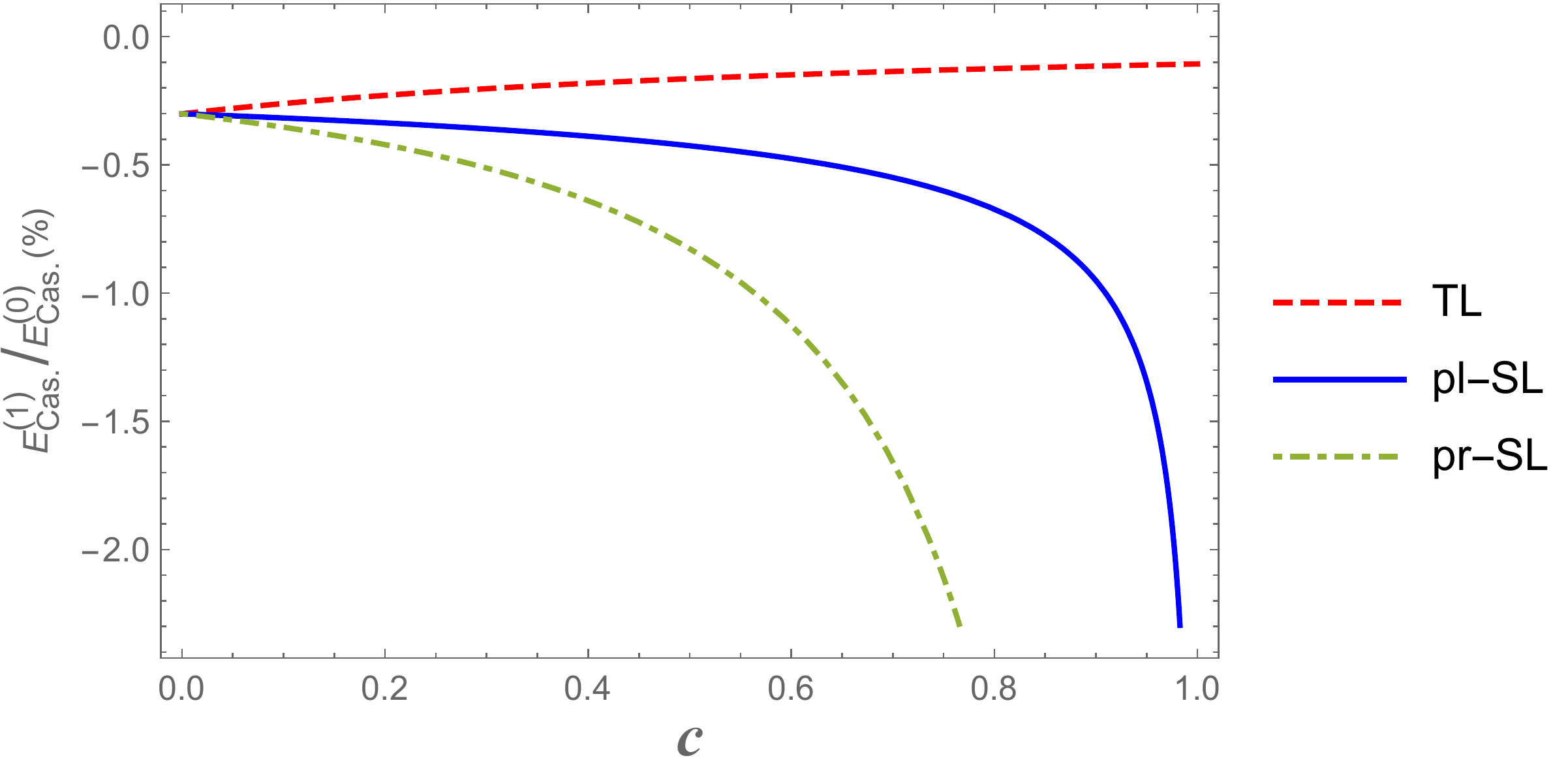}\caption{\label{fig.3} {\small The variation of the ratio between first order radiative correction and leading term, $E^{(1)}_{_{\mbox{\tiny Cas.}}}/E^{(0)}_{_{\mbox{\tiny Cas.}}}$, in terms of the Lorentz violating parameter $c$, with $\lambda=0.1$, $ m=1$ and $a=10 (\lambda_m)$. }}
		\label{fig2}
	\end{center}
\end{figure}
We can also compute the large mass and massless limits:
\begin{equation}
\left\{
\begin{array}{ll}
E^{(1)}_{_{\mbox{\tiny Cas.}}}\quad{\buildrel {am\gg1 } \over
	\longrightarrow }\quad
\displaystyle\frac{ 3 \ 
	L^2 }{1024 \pi^{7/2}}\frac{\lambda}{(1-c)^{1/2} a^3} \ (am)^{5/2} \ln(am)
\  e^{\frac{-2am}{\sqrt{1-c}},}  \\
\raisebox{-9mm}{$E^{(1)}_{_{\mbox{\tiny Cas.}}}
	\quad {\buildrel {m\to0 } \over
		\longrightarrow }\quad \displaystyle -\frac{L^2\lambda (1-c)^{1/2} }{512\pi^4 a^3}
	\left(\sum_{j=1}^{\infty}\frac{1}{j^2}\right)^2=-\frac{L^2\lambda (1-c)^{1/2}}{18432 a^3},$}

\end{array}\right.
\end{equation}
In figure \ref{fig1}, we have illustrated the variation of the ratio between the first order radiative corrections and leading terms, $E^{(1)}_{_{\mbox{\tiny Cas.}}}/E^{(0)}_{_{\mbox{\tiny Cas.}}}$, in terms of plates separation, for three distinct cases TL, pl-SL and pr-SL. We have also plotted this ratio in terms of Lorentz violating parameter $c$ in figure \ref{fig2}.

\section{Conclusion}
In this paper we have calculated the next to leading order radiative correction to the Casimir energy for $\phi^4$ theory with Lorentz-breaking symmetry in the context of renormalized perturbation theory. Our approach to calculate this energy is box renormalization method  introduced firstly by Boyer \cite{boyer1968quantum} and used for example in \cite{moazzemi2007dirichlet,Gousheh2009,Valuyan2008,Moazzemi2008}. The violation of symmetry breaking can be appeared in the Lagrangian by insertion of a term which couples the derivative of a field to a constant vector $u^\mu$. This additional term in the Lagrangian modifies the  dispersion relation and accordingly propagators of the fields. Therefore, in addition to the leading terms of physical quantities, all their sub-leading corrections are also affected. In  three separate cases of the Lorentz violation, violation in the time direction (TL), in the directions parallel (pl-SL) and perpendicular (pr-SL) to the plates, the leading terms of Casimir energy for $\phi^4$ theory have  been recently calculated in \cite{cruz2017casimir}. Here, we have investigated NLO corrections. We have plotted our results in figures \ref{fig1} and \ref{fig2}.

\end{document}